\documentclass[11pt]{cernrep}
\usepackage{graphicx}
\begin{document}

\title{The 2PI coupling expansion revisited}

\author{Anders Tranberg}

\institute{Department of Physics and Astronomy,\\University of Sussex,\\ Falmer, Brighton,\\ East Sussex BN1 9QH, UK.}

\maketitle

\begin{abstract}
Recently, out-of-equilibrium field theory has been studied using approximations based on truncations of the 2PI effective action. Although results are promising, the convergence of subsequent orders of the approximation is difficult to get a handle on, mainly because, generically, only the lowest non-trivial order is currently numerically tractable. We study one specific case, the broken phase of the $\phi^{4}$ model, where the existence of an effective three point vertex makes it possible to compare a lowest and next-to-lowest non-trivial order. 
\end{abstract}

%INTRODUCTION

\section{Introduction}
Out-of-equilibrium quantum fields can be treated using real-time equations of motion based on truncations of the 2PI effective action \cite{BerCox,Bergesreview} (based on the formalism of \cite{KadBay,CorJacTom}). The equations are written in terms of the (connected) propagator $\langle\hat{\phi}({\bf x},t)\hat{\phi}({\bf 0},t)\rangle_{c}=G({\bf x},t,t')=F({\bf x},t,t')-{\rm sign}(t-t')i\rho({\bf x},t,t')/2$, the mean field $\langle\hat{\phi}({\bf 0},t)\rangle=\phi(t)$ and the self-energies $\Sigma({\bf x},t,t')=\Sigma_{\rm loc}({\bf 0},t)\delta(t-t')\delta({\bf x})+\Sigma^{F}({\bf x},t,t')-{\rm sign}(t-t')i\Sigma^{\rho}({\bf x},t,t')$, and read
\begin{eqnarray}
\left[-\partial_{t}^{2}+\partial_{\bf x}^{2}+M^{2}_{\rm eff}(t)\right]F({\bf x},t,t')&=&\int_{0}^{t}dt''\int d^{3}{\bf z}\,\Sigma^{\rho}({\bf z},t,t'')F({\bf x-z},t'',t')\\
&-&\int_{0}^{t'}dt''\int d^{3}{\bf z}\,\Sigma^{F}({\bf z},t,t'')\rho({\bf x-z},t'',t'),\nonumber\\
\left[-\partial_{t}^{2}+\partial_{\bf x}^{2}+M^{2}_{\rm eff}(t)\right]\rho({\bf x},t,t')&=&\int_{t'}^{t}dt''\int d^{3}{\bf z}\,\Sigma^{\rho}({\bf z},t,t'')\rho({\bf x-z},t'',t'),\nonumber\\
\left[-\partial_{t}^{2}+M^{2}_{\rm eff}(t)-\frac{\lambda}{3}\phi_{t}^{3}\right]\phi_{t}&=&\frac{\delta \Phi[\phi,G]}{\delta\phi(t)}.
\label{eom}
\end{eqnarray}
with $M^{2}_{\rm eff}(t)=m^{2}+\Sigma_{\rm loc}({\bf 0},t)$. A $\Phi$-derivable approximation results from specifying a truncation of the infinite set of 2PI diagrams $\Phi$ contributing to the self energy through, 
\begin{equation}
\Sigma({\bf x},t,t')=-2\frac{\delta\Phi[\phi,G]}{\delta G({\bf x},t',t)}.
\end{equation}
Initial conditions are set by specifying an initial gaussian density matrix through 
\begin{equation}
\langle\{\hat{\phi}_{0}({\bf k}),\hat{\phi}_{0}({\bf -k})\}\rangle=(n_{k}+1/2)/\omega_{k},~~~ \langle\{\hat{\pi}_{0}({\bf k}),\hat{\pi}_{0}({\bf -k})\}\rangle=(n_{k}+1/2)\omega_{k},~~~\langle\{\hat{\pi}_{0}({\bf k}),\hat{\phi}_{0}({\bf -k}\})\rangle=0,
\end{equation}
for some choice of distribution function $n_{k}$ and dispersion relation $\omega_{k}$, and implementing the relations $\langle[\hat{\phi}_{0}({\bf k}),\hat{\phi}_{0}({\bf -k})]\rangle=0$, $\langle[\hat{\pi}_{0}({\bf k}),\hat{\phi}_{0}({\bf -k})]\rangle=1$.

Two expansions are on the market; one in the $O(N)$ model in powers of $1/N$ \cite{AarAhr,Bergesreview,Berges:2002cz,Arrizabalaga:2004iw} which has now become commonly used, and one in the coupling \cite{BerCox,Aarts:2001qa,JucCasGre}, which we will use here. Although it may potentially be less broadly applicable, it is straightforward to use and easy to understand. It is important to stress that the equations are self-consistent in a similar way to a gap equations, and that they resum infinite sets of diagrams.

%THIS WORK

\section{This work}
We study truncations of the coupling expansion of $\phi^{4}$ theory, with the action
\begin{equation}
S = -\int dt\,d^{3}{\bf x}\,\left(\frac{1}{2}\partial_{\mu}\hat{\phi}({\bf x},t)\partial^{\mu}\hat{\phi}({\bf x},t)+\frac{m^{2}}{2}\hat{\phi}^{2}({\bf x},t)+\frac{\lambda}{24}\hat{\phi}^{4}({\bf x},t)\right),
\end{equation}
\begin{figure}[t]
\begin{center}
\includegraphics[width=6in]{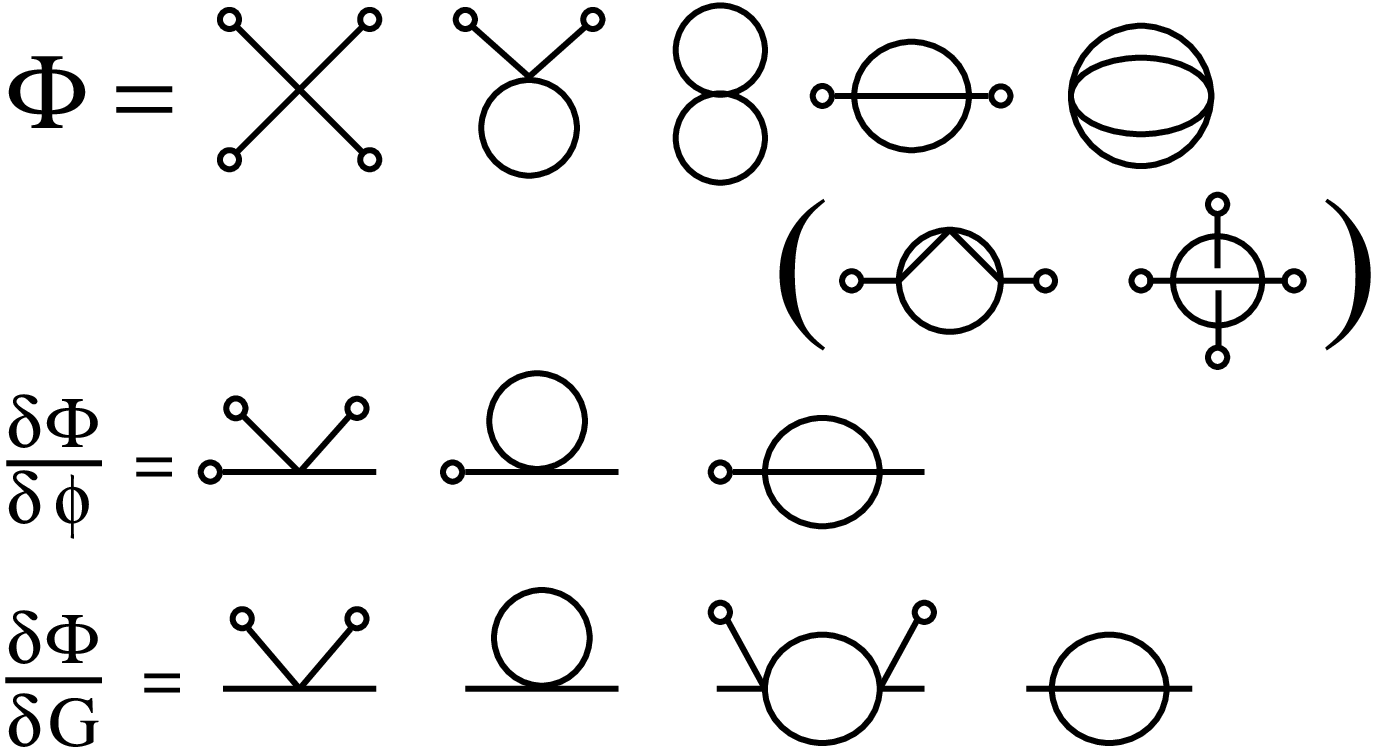}
\setcounter{figure}{1}
\caption{The included diagrams to $O(\lambda^{2})$ in the effective action (first line). Lines are full propagators $G$, small circles the mean field $\phi$. In the symmetric phase ($\phi=0$) the first, second and fourth diagram vanish. In the broken phase $\phi\propto\lambda^{-1/2}$, they are all there, as well as two more at $O(\lambda^{2})$ (in parenthesis), which we do not include . Below, the derived self-energy contributions for the mean field (second line) and propagator (third line). }
\label{fig1}
\end{center}
\end{figure}
The model has two phases: the ``symmetric'' ($m^2>0$), where $\langle\hat{\phi}\rangle=0$, and the ``broken'' ($m^2<0$), with $\langle\hat{\phi}\rangle=v=\sqrt{6\,m^2/\lambda}$. Figure 1 shows the diagrams at the level of the action and the resulting contributions to the self-energies in orders of $\lambda$ in either phase. The expansion is truncated at the order which is currently numerically tractable; including further vertices leads to additional nested space-time ``memory integrals'' on the right hand side of the equations of motion (\ref{eom}). At $O(\lambda^2)$, two additional diagrams exist for non-zero mean field. They are also not numerically tractable, and we consider the one (Basketball) diagram included here to represent the effect of the full $O(\lambda^2)$ contribution. The aim is to determine whether one order dominates the next.

In the symmetric phase, only the ``leaf'' diagram contributes at $O(\lambda)$, leading to the Hartree approximation. We will consider this a trivial approximation, since it is simply a (time-dependent) contribution to the effective mass (Re$\Sigma=\Sigma_{\rm loc}({\bf 0},t)$), and it is included in the quantity $M^{2}_{\rm eff}(t)$. Mode damping, equilibration and thermalization, processes of interest in out-of-equilibrium field theory, are the result of a non-zero {\it imaginary} part of the self-energy and in the symmetric phase, this appears at $O(\lambda^2)$, in the last (Basketball) diagram. Detailed study of thermalization and equilibration in this truncation in 1, 2 and 3+1 dimensions can be found  in \cite{BerCox,JucCasGre,ArrSmiTra}. In particular, the damping rate can be calculated in perturbation theory, the result being qualitatively compatible with the 2PI resummed result \cite{ArrSmiTra}. Apparently, the resummation has only a small, quantitative effect in this case.

In the broken phase, the mean field leads to an effective three point vertex, and because $\langle\hat{\phi}\rangle$ is formally $O(\lambda^{-1/2})$, an additional 2-loop diagram appears at order $O(\lambda)$ (Two-loop) in the effective action. Although in perturbation theory it does not lead to on-shell damping \cite{ArrSmiTra}, resummation from the self-consistent evolution equations leads to a non-zero Im$\Sigma$. In this case, we are able to compare two successive truncations of Im$\Sigma$ and estimate the quality of the expansion.

%RESULTS

\section{Results}
Figure 2 shows the evolution of the effective particle number $n_{k}+1/2=\sqrt{F({\bf k},t,t')\partial_{t}\partial_{t'}F({\bf k},t,t')}_{t=t'}$ in time. We start with a ``top-hat'' (T1) initial condition which is very out-of-equilibrium, and we can see a progression to a smooth Bose-Einstein-like distribution at time $mt\simeq 500$. The dark (black) dots are $O(\lambda^2)$, the light (green) $O(\lambda)$. It is clear, that not only does the resummation lead to a quantitative effect (the non-zero Im$\Sigma$ at $O(\lambda)$), but the time scales of equilibration are also very similar. Here, $\lambda=1$. This constitutes {\it kinetic} equilibration, and will (at small coupling) lead to a Bose-Einstein distribution with an effective temperature and chemical potential. 
\begin{figure}[t]
\begin{center}
\includegraphics[width=6in]{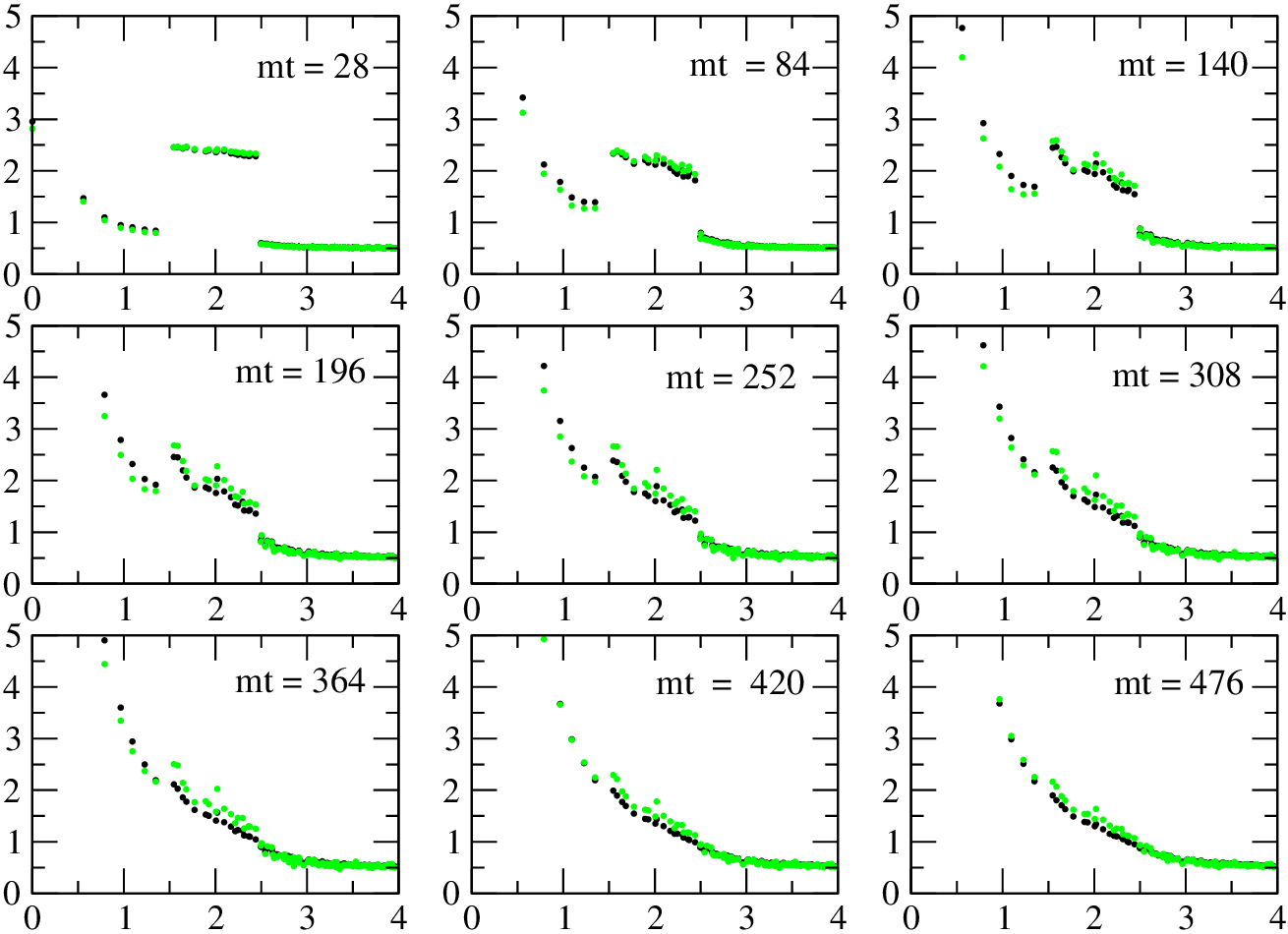}
\setcounter{figure}{2}
\caption{Distribution function $n_{k}+1/2$ at various times. Around time $mt=500$, the curve is smooth and approaching a Bose-Einstein distribution.}
\label{fig2}
\end{center}
\end{figure}
The $\phi^{4}$ model has no conserved charges, and the asymptotic equilibrium state has zero chemical potential. To reach it requires {\it chemical} equilibration to take place as well. 

Figure 3 shows $\ln(1+1/n_{k})$ vs. $\omega_{k}=\sqrt{\partial_{t}\partial_{t'}F({\bf k},t,t')/F({\bf k},t,t')}_{t=t'}$ at $mt=1000$. A Bose-Einstein distribution is then a straight line with slope $T^{-1}$ and intercept $\mu_{\rm ch}/T$. Overlaid are the results for Bose-Einstein distributions with zero chemical potential. Clearly, the low-momentum range has equilibrated kinetically, but not (yet) chemically. It is well known that at $O(\lambda^2)$, chemical equilibration takes place (see for instance \cite{Berges:2002wr,JucCasGre,ArrSmiTra}). In figure 4 we plot the total particle number $N_{\rm tot}=\int d^{3}{\bf k}\, n_{k}$ normalized to the initial value. In both approximations, there is a change of particle number. Although the behaviour is not the same, presumably they will both lead to chemical equilibration.
\begin{figure}[t]
\begin{center}
\includegraphics[width=4in]{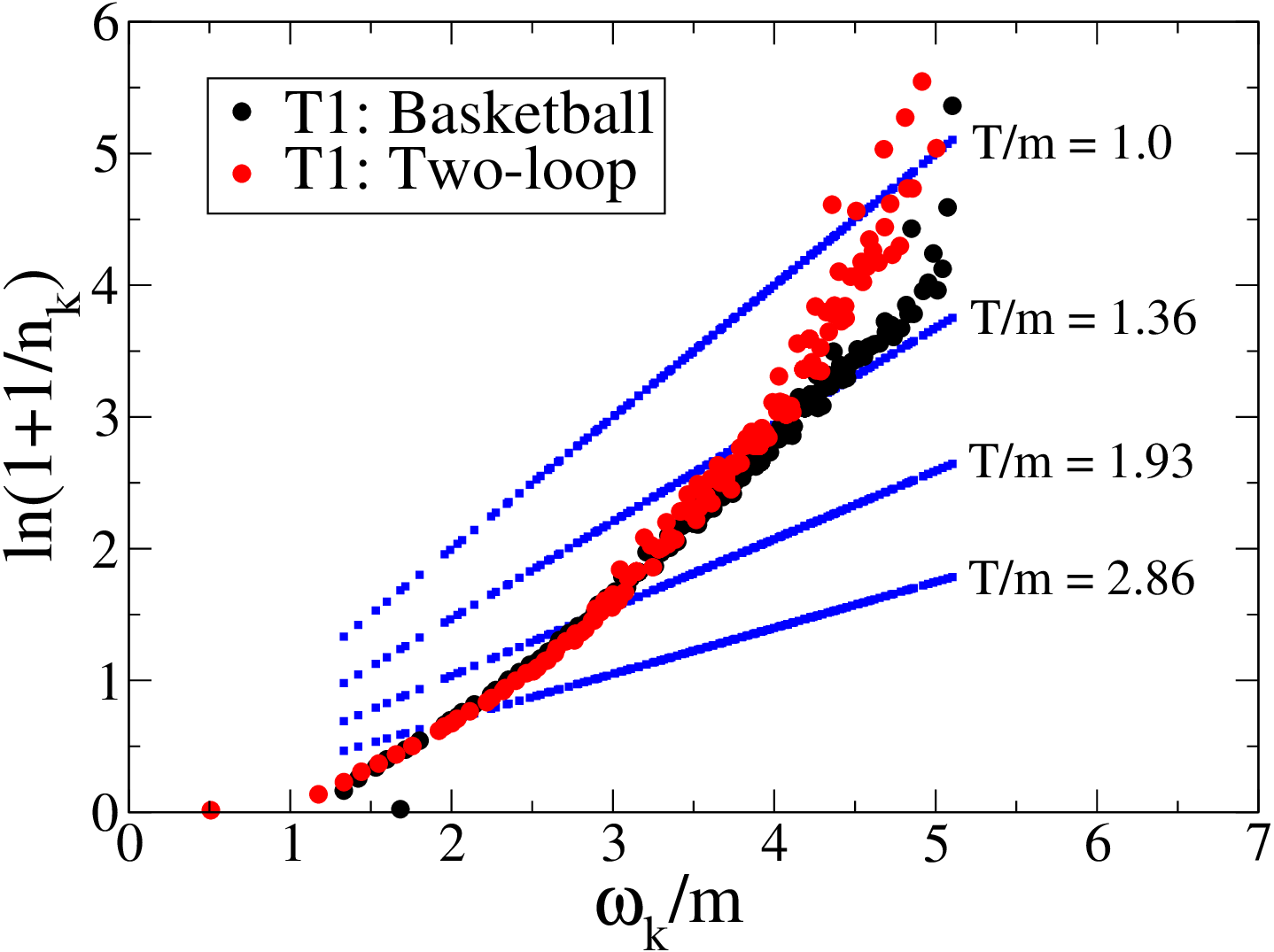}
\setcounter{figure}{3}
\caption{The distribution function $\ln(1+1/n_{k})$ at time $mt=1000$. The low momentum range has kinetically equilibrated. The straight line are Bose-Einstein distributions at various temperatures and zero chemical potential.}
\label{fig3}
\end{center}
\end{figure}

\begin{figure}[t]
\begin{center}
\includegraphics[width=4in]{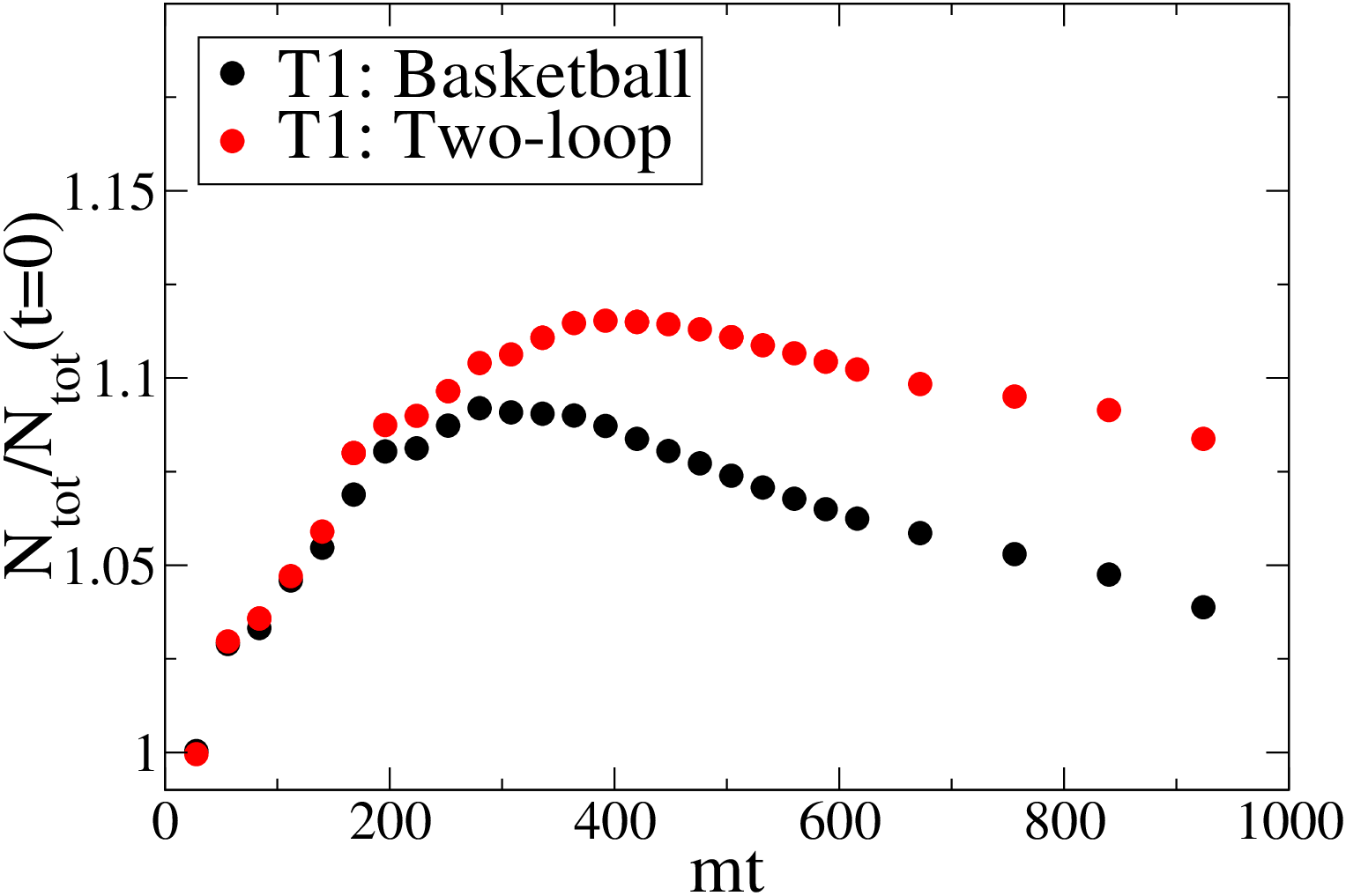}
\setcounter{figure}{4}
\caption{The total particle number in time.}
\label{fig4}
\end{center}
\end{figure}

%CONCLUSION

\section{Conclusions}
Successive, non-trivial truncations of the coupling expansion of the 2PI/$\Phi$-derivable approximation have been compared. In the broken phase of $\phi^4$ theory, kinetic and chemical equilibration is included at $O(\lambda)$, at a rate very close to what we find at $O(\lambda)+O(\lambda^{2})$. At this moderate coupling, the expansion seems to be well behaved. In addition, we found that the relevant diagram at $O(\lambda)$, which perturbatively does not lead to on-shell damping, becomes the dominant contribution when resummed by the self-consistent evolution equations. 

Further investigation of the convergence of expansions of the 2PI effective action is needed. Because of the numerical effort required, it may be necessary to consider simpler systems, such as quantum mechanics, as a testing ground. For further details and a study of equilibration in the symmetric phase, see \cite{ArrSmiTra}.

%ACKNOWLEDGMENTS

\section*{Acknowledgments}
I would like to thank Jan Smit and Alejandro Arrizabalaga for collaboration on this work, Gert Aarts for fruitful discussions and the organizers for an enjoyable meeting. This work is supported by FOM/NWO and the PPARC SPG ``Classical Lattice Field Theory''.

%BIBLIOGRAPHY

\end{document}